\documentclass[prl,aps,twocolumn,showpacs]{revtex4}
\usepackage{amsmath}
\usepackage{amssymb} \usepackage{graphicx}

\begin{document}
\title{Relaxation in time-dependent current-density
  functional theory}
\author{Roberto D'Agosta}
\affiliation{Department of Physics and Astronomy, University of
Missouri, Columbia, Missouri 65211}
\author{Giovanni Vignale}
\affiliation{Department of Physics and Astronomy, University of
Missouri, Columbia, Missouri 65211}
\date{\today}

\begin{abstract}
We apply the time-dependent current-density functional theory to the
study of the relaxation of a closed many-electron system evolving from
an non-equilibrium initial state. We show that the self-consistent
unitary time evolution generated by the time dependent
exchange-correlation vector potential irreversibly drives the system
to equilibrium. We also show that the energy dissipated in the
Kohn-Sham system is related to the entropy production in the real
system. 
\end{abstract}
\maketitle

The irreversible time evolution of a many-electron system which is
initially prepared in a non-equilibrium state is the subject of strong
research interest \cite{Weiss1999}. The system might be, for example,
a large molecule \cite{Cuniberti2005} or the electron liquid in a
quantum well, and the initial non-equilibrium state might be created
by laser excitation, or by the application and subsequent removal of
an electric field \cite{Gavrila1992}. From experience we know that
even if the system is closed and isolated, i.e. no exchanges of
particles and energy are allowed, the interaction between the
particles drives the system towards the state of maximum entropy,
i.e., the equilibrium state. When this happens we say that the system
has relaxed \footnote{We assume that the system is large enough for
the Poincar\'e's recurrence time to be pratically infinite.}. At each
stage of the relaxation process the system has a certain amount of
``mechanical energy" which could in principle be channeled into work. 
One of the tasks of theory is to develop efficient tools
--in alternative to the solution of the many-body Schr\"odinger
equation-- to calculate the time evolution of the mechanical energy. 
One would like to do this starting from first principles, yet without
knowing the exact state of the many-body system as a function of time.

Various formulations of the problem of relaxation of a quantum system
have been proposed over the years. Most of these formulations couple
the system of interest to some kind of external environment --the
so-called thermal bath \cite{Feynman1963}. When the information about
the microscopic state of the thermal bath is discarded this approach
leads to an effective dynamics of the system that is no longer
unitary. The main weakness  of this approach lies in the fact that the
thermal bath and its coupling to the system must be modeled in an
essentially ad-hoc manner.   In this paper we will describe an
alternative approach, based on the time-dependent current-density
functional theory (TDCDFT)
\cite{Hohenberg1964,Kohn1965,Gross1996,
vanLeeuwen2001,Giulianivignale}, which allows a first-principle
treatment of relaxation in terms of a self-consistent unitary time
evolution \footnote{For a different approach to dissipative systems in
the framework of time-dependent density-functional theory, see K.
Burke, R. Car, and R. Gebauer, Phys. Rev. Lett. {\bf 94}, 146803
(2005).}.

We assume that the external potential acting on the system does not
depend on time and that the system is closed and isolated. We thus
consider the time evolution of a many-electron system described by the
Hamiltonian 
\begin{equation}
\hat H=\sum_i \frac{\hat p_i^2}{2m}+\hat W+\int dr~
V(r)\hat n(r),
\end{equation}
where $\hat n(r)$ is the density operator, $V(r)$ is the external
potential and $\hat W$ describes the electron-electron interaction
(here and in the following $\hbar=1$). A possible way to generate the
initial non-equilibrium state $|\psi_i\rangle$ is by applying an
external potential at $t=-\infty$ and allowing the system to relax in
the presence of this potential. The external potential is then
switched off at the initial time $t=0$ and the system evolves
henceforth  according to the Schr\"odinger equation $i\partial_t
|\Psi(t)\rangle =\hat H |\Psi(t)\rangle$. Due to the presence of the
electron-electron interaction the system will relax to an equilibrium
state of $\hat H$. Because the system is closed and isolated during
its time evolution, the energy is conserved and equal to its initial
value $E_i=\langle\psi_i|\hat H|\psi_i\rangle$. This implies that the
final equilibrium state will not be the ground state of $\hat H$, but
the most probable state with total energy $E_i$. In this final state
the entropy, as well as the temperature will have finite values. Our
task is to calculate the time evolution of the mechanical energy
between the initial and the final state of this evolution.

The central tenet of TDCDFT is that the correct time evolution of the
particle and current densities of the many-electron system  can be
obtained from the time evolution of a {\it non-interacting} reference
system, known as the Kohn-Sham (KS) system, described by the state
$|\Psi_{KS}(t)\rangle$. This state evolves from an initial state
$|\Psi_{KS,i}\rangle$ --constructed to give the initial density and
energy of the system-- according to the equation
\begin{equation} \label{KS.equation}
i\partial_t|\Psi_{KS}(t)\rangle=\hat H_{KS}(t)|\Psi_{KS}(t)\rangle
\end{equation}
 where the KS Hamiltonian is
\begin{equation}\label{HKS}
\begin{split}
\hat{H}_{KS}(t) =&\sum_i
\frac{1}{2m}\left[\hat{p}_i+eA_{xc}(\hat
  r_i,t)\right]^2\\ &+ \int dr~ \hat
n(r)V_{hxc}(r,t)+\int dr~\hat n(r)V(r).
\end{split}
\end{equation}
$V_{hxc}(r,t)$  is an effective scalar potential that includes the
Hartree potential (h) as well as the ground-state exchange-correlation
(xc) potential of static density functional theory (DFT) evaluated at
the instantaneous density $n(r,t)$.  $A_{xc}(r,t)$ is the xc  vector
potential, which is a functional of the density and of
the time-dependent current density $\vec j (r,t)$, and whose role is
to enforce the correct time evolution of the current. The time
dependence of $\hat H_{KS}$ arises from its being a functional of $n$
and $\vec j$. In practice, both  $V_{hxc}(r,t)$  and $A_{xc}(r,t)$
need to be approximated, but these approximations are non-empirical,
in the sense that they are based on exact constraints coming from the
physics of the homogeneous electron liquid \cite{Giulianivignale}. For
$V_{hxc}(r,t)$ the simplest and most popular approximation is the
local density approximation (LDA), which expresses the xc potential as
a functional of the density in the following manner:
\begin{equation}\label{LDA}
V_{hxc}(r,t)=\int dr'~\frac{e^2}{|r-r'|} n(r',t)+\mu_{xc}(n(r,t))~,
\end{equation}
where $\mu_{xc}(n)=d \epsilon_{xc}(n)/dn$ and $\epsilon_{xc}(n)$ is
the xc energy density of a uniform electron gas
\cite{Giulianivignale}.

As for $A_{xc}$, a quasi-local approximation in terms of the
current density was derived by Vignale and Kohn
\cite{Vignale1996}, and can be cast in  the form \cite{Vignale1997b}
\begin{equation}
\label{axc}
e\frac{d}{dt}{A}_{xc,k}(r,t)= \frac{1}{n(r,t)}\nabla_i
\sigma_{xc}^{ik}(r,t)
\end{equation}
where $i,k$ denote cartesian components and the stress tensor
$\sigma_{xc}$ is given by
\begin{equation}
\begin{split}
\sigma_{xc}^{ik}(r,t)=& 
\zeta(r,t)\delta_{ik}\nabla\cdot v\phantom{\frac12}\\
&+\eta(r,t)\left(\nabla_i v_k+\nabla_k
  v_i-\frac{2}{3}\delta_{ik}\nabla\cdot
  v\right)
\end{split}
\label{sigmaxc}
\end{equation}
where $v(r,t)=j(r,t)/n(r,t)$ is the velocity, and we have, for
simplicity, neglected retardation effects in the relation between
$\sigma_{xc}$ and the velocity. The quantities $\eta$ and $\zeta$ are
the visco-elastic constants of the homogeneous electron liquid and can
be calculated from the linear response functions of the
latter \cite{Giulianivignale}, evaluated at the instantaneous density
$n(r,t)$.

We want to point out that the Eqs.~(\ref{KS.equation})-(\ref{sigmaxc})
define a self-consistent unitary evolution which is designed to
approximate the true density and current of the system.  It is the
dependence of the xc vector potential on the velocity that
breaks the time-reversal invariance of the KS hamiltonian and allows
for the possibility of irreversible relaxation.

We now define the key quantity of this paper, the  ``Kohn-Sham energy"
\begin{equation}
\begin{split}
E(t)=&\langle\Psi_{KS}(t)|\hat
H_{KS}(t)|\Psi_{KS}(t)\rangle\\
&-\int dr~V
_{hxc}(r,t)n(r,t)+E_{hxc}[n(t)]
\label{ksenergy}
\end{split}
\end{equation}
where $E_{hxc}$ is the Hartree + xc energy functional of the
ground-state DFT.  $E_{hxc}$  and $V_{hxc}$ are related by the
identity  $V_{hxc}=\delta E_{hxc}(n)/\delta n$. In static DFT, $E(t)$
is just the familiar expression from which the ground state energy is
calculated. In TDCDFT, $E(t)$ is not the true energy of the
interacting
many-particle system, but, as we will show below, it decreases
monotonically with time going from $E_i$ at $t=0$ to some final value
$E_f\ge E_0$  at $t = +\infty$, $E_0$ being the true ground-state
energy of the system.

To prove this point we observe that a straightforward
calculation shows that (we used $\delta \hat H_{KS}/\delta
A_{xc}=\hat j$)
\begin{equation}
\frac{dE}{dt}=\int dr~ j(r,t)\cdot \frac{dA_{xc}}{dt}.
\label{currentdissipation}
\end{equation}
Eq.~(\ref{currentdissipation}) immediately allows us to identify
$dE/dt$ as the work done on the system by the self-consistent xc
vector potential. By putting Eq.~(\ref{axc}) in
Eq.~(\ref{currentdissipation}) we get
\begin{equation}
\label{dedt}
\frac{dE}{dt}=-2\int dr~\eta \mathrm{Tr}\left(\omega-\frac13
\mathrm{Tr} \omega\right)^2-\int dr~\zeta (\mathrm{Tr}
\omega)^2,
\end{equation}
where, $\omega_{ij}=(\nabla_i v_j+\nabla_j v_i)/2$.
The non-positivity of $dE(t)/dt$ follows from
the fact that the viscosity constants $\eta$ and $\zeta$ are
positive \cite{Giulianivignale}.

Next we prove that $E(t) \to E_f\geq E_0$ for $t \to \infty$.
We begin by observing that, if
$\hat T_{A_{xc}}=\sum_{i}(\hat p_i+eA_{xc})^2/2m$ is the KS kinetic
operator, then at a fixed time $t$
\footnote{We have $E(t)=\langle
 \Psi_{KS}(t)|\hat T_{A_{xc}}+\hat
V|\Psi_{KS}(t)\rangle+E_{hxc}[n(t)]$ and
$E_0[A,V]=\min_{\Psi}\{\langle \Psi|\hat T_{A}+\hat
V|\Psi\rangle+E_{hxc}[n,A]\}$.},
\begin{eqnarray}\label{inequality.1}
E(t)&\ge& \min_{\Psi\to n(t)}\langle
 \Psi(t)|\hat T_{A_{xc}(t)}+\hat V|\Psi(t)\rangle+E_{hxc}[n(t)]\nonumber\\
&\ge& E_0[A_{xc}(t),V]+E_{hxc}[n(t)]-E_{hxc}[n(t),A_{xc}(t)]\nonumber\\
\end{eqnarray}
where $E_0[A,V]$ is the instantaneous ground state energy of the system, in
the presence  of static potentials $V$ and $A$. $E_{hxc}[n,A]$
is the Hartree+xc energy functional of the system in the presence of a
static vector potential $A(r)$. Now it is known \cite{Vignale1987}
that in LDA and up to second-order in
$B(r)\equiv \nabla \times A(r)$, one has  $E_{hxc}[n]-E_{hxc}[n,A] =
\int dr~c(n(r)) B^2(r) >0$ because $c(n)$ ($=$ one half the difference
between the orbital magnetic susceptibilities of the interacting and
non-interacting homogeneous electron liquid \cite{Vignale1987})  is a
positive quantity.
Assuming that this inequality is generally true we can then neglect
the last two terms on the right hand side of Eq.~(\ref{inequality.1}) 
to arrive at $E(t)\ge E_0[A_{xc}(t),V]$. We now observe that, because
the electron-electron Coulomb interaction is a positive operator, the
ground-state energy of an interacting many-particle system is always
greater than the ground-state energy of the same system where
electron-electron interaction is turned off. We then conclude that
$E_0[A_{xc},V]\ge E_0^{ni}[A_{xc},V]$ where $E_0^{ni}$ is the
instantaneous ground state energy of the non-interacting system. On
the other hand, the ground state energy of a non-interacting fermion
system is always greater than the ground state energy, $E^{ni}_{0,b}$,
of a non-interacting boson system in the same external potentials. We
then have
\begin{equation}
E(t)\ge E_0^{ni}[A_{xc},V]\ge E^{ni}_{0,b}[A_{xc},V]\ge E^{ni}_{0,b}[A_{xc}=0,V],
\label{inequality}
\end{equation}
where we have used in the last step the diamagnetic inequality, which
states that the ground state energy of a spinless boson system in the
presence of a vector potential is always greater than the ground state
energy of the same system without the vector potential
\cite{Simon1976}. Because $E^{ni}_{0,b}[A_{xc}=0,V]$ is a
time-independent quantity we have arrived at our conclusion that
$E(t)$ is bounded. Eqs. (\ref{dedt}) and (\ref{inequality}) allow us
to conclude that the limit $t\to \infty$ of $E(t)$ is finite and that
$\lim_{t\to \infty}dE/dt=0$. From Eq.~(\ref{dedt}) and the requirement
that $j$ vanishes at infinity, we deduce that we can have $dE/dt=0$ if
and only if $\nabla \cdot v=0$ and $\nabla\times v=0$, which imply
$v=0$ everywhere (we consider only simply connected geometries). In
turn this implies that the current $j$ is zero everywhere, and
therefore, through the continuity equation, the density approaches a
stationary limit $n_f(r)$. This can only happen if the asymptotic
state is one of the eigenstates of the KS Hamiltonian with density
$n_f(r)$. In the following, we will disregard the unlikely cases when
the system gets stuck in some excited state,
and we will assume $n_f(r)=n_0(r)$, where $n_0( r)$ is the
ground-state density.  The time evolution of Eq.~(\ref{KS.equation}) 
will then drive the KS system to the ground-state of the corresponding
KS hamiltonian.

\begin{figure}[t]
\includegraphics[width=7cm,clip]{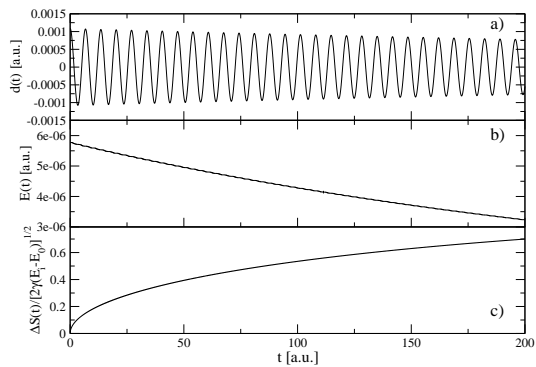}
\caption{a) Plot of the dipole moment $d(t)=\int dzz n(z,t)
$. b) Plot of the Kohn-Sham energy $E(t)-E_0$. c) Plot of
$\Delta S(t)/\sqrt{2\gamma (E_i-E_0)}$ as given by
Eqs.~(\ref{entropyexample}) and (\ref{exponentiale}). A
fit with $E(t)$ in Eq.~(\ref{exponentiale}) of
the data in plot b) gives $\Gamma=0.0034$ a.u. We have always
used $\mathcal{E}=0.01$ mV/nm.}
\label{energies-plot}
\end{figure}
To demonstrate the above ideas in a concrete example, we have computed numerically the time evolution of
the electronic dipole moment $d(t)=\int dz z n(z,t)$
(Fig.~\ref{energies-plot}, panel a) and the
energy $E(t)-E_0$ (Fig.~\ref{energies-plot}, panel b)  in a
one-dimensional quantum well, such
that the electrons are confined in the $z$-direction, and yet are free
to move in the $x-y$ plane \cite{Wijewardane2005}. The single-particle
states for this
system are of the form $\psi_n(z) e^{i  k \cdot  r}$ where
$k$ and $r$ are vectors in the $x-y$ plane. The system is initially
prepared in a state that corresponds to the KS ground-state in the
presence of a uniform electric field  $\mathcal{E}=0.01$ mV/nm in
the $z$-direction. This state is homogeneous in the $x-y$ plane, and
this property is maintained throughout the subsequent time evolution.
This means that all the electrons share the same state of motion in
the $z$-direction --a time-dependent state that can be written as a
linear superposition of the lowest-lying stationary states $\psi_n(z)$
(typically the two lowest ones suffice). The xc vector potential
$A_{xc}$ has been approximated in the form of Eq.~(\ref{axc})
\footnote{A more accurate study of this system, including the
frequency-dependence of the visco-elastic coefficients has been
recently presented in Ref.~\onlinecite{Wijewardane2005}.}. We see from
Fig.~1 that the condition $dE/dt\le 0$ is verified at all times. 
Moreover the energy $E(t)$ is well approximated by the analytic form
\begin{equation}
\label{exponentiale}
E(t)-E_0=(E_i-E_0)e^{-\Gamma t},
\end{equation}
where the parameter $\Gamma$ can be estimated from the slope of the
curve in Fig. \ref{energies-plot}.

\begin{figure}[t]
\includegraphics[height=4.5cm,clip]{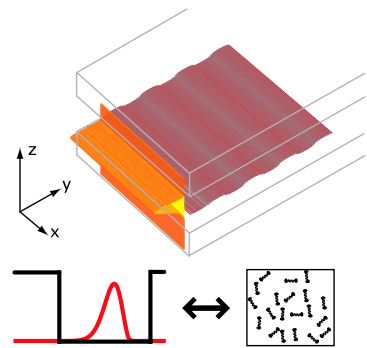}
\caption{The motion in the $z$-direction is coupled through the
Coulomb interaction to a 2DEG in the $x-y$ plane. The energy is lost
to the motion in the $z$-direction at a rate given by $dE/dt$.}
\label{twosystems}
\end{figure}

We now wish to argue that $E(t)-E_0$ is the maximum work that can be
extracted from the system at a given time, and that its time
derivative is related to the rate of entropy production \cite{Landau5}. To do
this, let us take a closer look to our quantum well system (see
Fig.~\ref{twosystems}).
In this system we can recognize a macroscopic degree of freedom
--the $z$-dependent wave function in which all the electrons reside-- 
and a ``thermal bath" --the two-dimensional electron-gas (2DEG)
in the $x-y$ plane.   The coupling between the two is induced by the
Coulomb interaction: no empirical modeling is needed. As the
macroscopic state of motion in the $z$-direction evolves with time,
energy is transferred from this motion into
low-lying excitations of the 2DEG in the $x-y$ plane.  These
excitations are, in the simplest case, double electron-hole pairs with
zero total momentum.  Let us study in detail this energy exchange.

We start from an initial state in which the 2DEG is in the
ground-state but all the electrons occupy an excited state of motion
in the $z$-direction.  At the end of the relaxation all the electrons
are in lowest energy state in the $z$-direction, but the 2DEG is
in an excited state.   The total energy is conserved and given by
$E_i$. On the other hand, the infinite-time limit of $E(t)$ is the
ground-state energy of the system, $E_0$, as one can see by noting
that the final Kohn-Sham hamiltonian, according to Eq.~(\ref{HKS})
coincides with the Kohn-Sham hamiltonian of static DFT which, by
definition, yields the correct ground-state energy.   Since the
electrons are finally at rest in the $z$-direction,  the difference
$E_i-E_0$ must be entirely ascribed to the excitation of the 2DEG.
This means that $E_i-E_0$ is the total heat, $Q$, transferred from the
macroscopic motion in the $z$-direction to the 2DEG.

Denoting by $T_i$ and $T_f$ the initial and final
temperature of the 2DEG  we have
\begin{equation}\label{heat}
Q=\int_{T_i}^{T_f} dT~C_V(T)~.
\end{equation}
where $C_V(T)$ is the heat capacity of the 2DEG.  At low temperature
we have $C_V (T)=\gamma T$ with $\gamma=\pi {\cal A} m^* k_B/3$ where
${\cal A}$ is the area occupied by the gas in the $x-y$ plane and $m^*$ is the particle
mass renormalized by electron-electron interactions. Eq.~(\ref{heat})
provides a quantitative relation between the increase in temperature
of the ``thermal bath" and the energy  that is lost to the 
macroscopic motion in the $z$-direction. Because in our model
calculation $T_i=0$ we get for the final temperature
$T_f=\sqrt{\frac{2}{\gamma}(E_i-E_0)}$
and by recalling that $C_V/T=dS/dT$ we get the increase in entropy
$\Delta S=\gamma T_f$.

The fact that the energy $E_i-E_0$ is converted into heat that
increases the temperature and entropy of the 2DEG allows us to
identify $E_i-E_0$ as the maximum energy that could in principle be
extracted from the motion in the $z$-direction to do work \footnote{In
the motion in the $z$-direction the entropy is constant.}. In other
words, $E_i-E_0$ is the work we would extract from the system if we
could prevent excitation of the 2DEG during the relaxation of the
motion in the $z$-direction. Any excitation of the 2DEG will
necessarily result in less work obtained from the system.

We can also use Eq.~(\ref{heat}) to define  a suitable ``temperature"
for non-equilibrium states. We begin with the observation that
Eq.~(\ref{heat}) can be cast in the form
\footnote{We have $E_0-E_i=\int_0^\infty
dt~dE/dt$.}
\begin{equation}
\int_0^\infty dt~\left[\frac{dE}{dt}+T(t)\frac{dS}{dt}\right]=0,
\label{conservation}
\end{equation}
 which simply states the global conservation of energy.
This suggest that we {\it define} the non-equilibrium temperature of the system in
such a way that the equation
\begin{equation}\label{deftemp}
\frac{dE}{dt}=-T(t)\frac{dS}{dt}
\end{equation}
is satisfied at all $t$.
The possibility of this definition stems from  the fact that
we can always follow a non-equilibrium thermodynamical
transformation through a series of small quasi-equilibrium
transformations. Then Eq.~(\ref{deftemp}) is sufficient for the
Eq.~(\ref{conservation}) to be satisfied.
Eq.~(\ref{deftemp}) is in fact a differential
equation \footnote{Observe that from $C_V/T=dS/dT$ follows $TdS/dt=C
_V dT/dt$.} for the temperature that can be easily solved to give
$T(t)=\sqrt{\frac{2}{\gamma}[E_i-E(t)]}$
(we have used again $C_V(T)=\gamma T$ for a
2DEG)  and finally (see Fig.~\ref{energies-plot}, panel c)
\begin{equation}
\Delta S(t)=\sqrt{2 \gamma [E_i-E(t)]}.
\label{entropyexample}
\end{equation}
This is the desired connection between $E(t)$ and $\Delta S(t)$.

In conclusion we have shown that the time-dependent current-density
functional theory naturally opens the possibility of a first-principle
description of relaxation in many-body systems. To show this we have
identified a quantity, the Kohn-Sham
energy, that defines a ``time arrow" in the evolution of the system.
Indeed we have proved that this energy monotically decreases with time
until an equilibrium state is reached. We are also able to relate the
energy dissipated to the increase in entropy and temperature. This
thermodynamical approach allows us to identify the Kohn-Sham energy
with the maximum amount of work that can be extracted from the
system.

\acknowledgments
We thank Carsten Ullrich for assistance  with the
numerical calculations, and for many useful
discussions and Janmin Tao for a critical reading of the
manuscript. We acknowledge financial support from NSF Grant No.
DMR-0313681 and the kind hospitality of the Scuola Normale
Superiore in Pisa where part of this work was completed.
\bibliography{dft-biblio,general}
\end{document}